\documentclass[10pt,conference]{IEEEtran}
\IEEEoverridecommandlockouts
\usepackage{cite}
\usepackage{amsmath,amssymb,amsfonts,amsthm}
\usepackage[]{algorithm2e}
\usepackage{graphicx}
\usepackage{textcomp}
\usepackage{subcaption}
\usepackage{xcolor}
\usepackage{multirow}
\usepackage{array}

\theoremstyle{definition}
\newtheorem{exmp}{Example}

\newcommand{\mx}[1]{\mbox{\boldmath $#1$}}

\newcommand{\bwa}{bandwidth allocation }

\newcommand{\BD}{B\! D}
\newcommand{\fs}{f\! s}

\newcommand{\TODO}[1]{\textcolor{red}{#1}}

\newcommand{\IGNORE}[1]{}

\def\BibTeX{{\rm B\kern-.05em{\sc i\kern-.025em b}\kern-.08em
    T\kern-.1667em\lower.7ex\hbox{E}\kern-.125emX}}
\begin{document}

\title{Multi timescale bandwidth profile and its application for burst-aware
fairness\\ 
\thanks{Partially supported by the OTKA grant K123914.}
}

\author{
	\IEEEauthorblockN{
		Szilveszter N\'adas\IEEEauthorrefmark{1},
		Bal\'azs Varga\IEEEauthorrefmark{1},
		Ill\'es Horv\'ath\IEEEauthorrefmark{2}, 
		Andr\'as M\'esz\'aros\IEEEauthorrefmark{3}, 
		Mikl\'os  Telek\IEEEauthorrefmark{3}
	}
	\IEEEauthorblockA{
		\IEEEauthorrefmark{1}Traffic Analysis and Network Performance Laboratory,
		Ericsson Research, Hungary\\
		\{szilveszter.nadas,balazs.a.varga\}@ericsson.com\\
		\IEEEauthorrefmark{2}MTA-BME Information Systems Research Group, Hungary\\ horvath.illes.antal@gmail.com\\
		\IEEEauthorrefmark{3}Dept. of Networked Systems and Services, Budapest University of Technology and Economics, Hungary\\
		\{meszarosa, telek\}@hit.bme.hu}
}

\sloppy

\maketitle

\begin{abstract}
We propose a resource sharing scheme that takes into account the
traffic history over several predefined time scales and provides fair resource sharing considering the traffic history. 
Our concept builds on a simplified version of core-stateless resource sharing,
where we only use a few Drop Precedences (DPs).
For packet marking we introduce Multi timescale bandwidth profile.
Additionally, we provide basic dimensioning concepts for the proposed schema
and present its simulation based performance analysis.
\end{abstract}


\section{Introduction}
Quality of Service (QoS) is a fundamental area of networking research that has
been researched for a long time.
Despite this several open issues remain as collected in \cite{RFC6077}. 
Per node (e.g.\ per subscriber or per traffic aggregate) fairness, being one of
these, is usually provided by per node WFQ, but that does not scale as the number
of node increases.
Core-stateless schedulers \cite{PPV,SCDemo18, abc} solve this, but they still
only provide fairness on a single timescale, typically at the timescale of
round-trip time (RTT).
These solutions mark packets per node at the edge of the network and do
simple, session-unaware scheduling in the core of the network based on the
marking.
The most common way to provide fairness on longer timescales is to
introduce caps, e.g.\ a monthly cap on traffic volume,
however the congestion lasts much shorter time period.
A similar attempt is to limit the congestion volume instead of the traffic
volume as described in \cite{RFCConEx,MobileConEx}.
The need for fairness on different time scales is illustrated by the example of
short bursty flows and long flows, as mice and elephants in \cite{Iyengar}.
A demand in this area is that a continuously transmitting node shall achieve the
same long term average throughput as nodes with small/medium bursts now and
then.

Traffic at the same time is becoming more and more bursty, including traffic
aggregates, due to the highly increased throughput of 5G Base
Stations \cite{eri20165g}. 
When deploying mobile networks, operators often lease transport lines as
Mobile Backhaul from the Core Network to the Base Stations. 
The transport services and related bandwidth profiles defined by the Metro
Ethernet Forum (MEF) \cite{MEF62} are most commonly used for this purpose
today.
However using the current service definitions,
it is not possible to achieve per node (per transport service in this case)
fairness and good utilization simultaneously and also it only takes into
account a small timescale, i.e.\ the instantaneous behavior.

In this paper we propose a resource sharing scheme that takes into account the
traffic history of nodes over several predefined timescales and
provides fair resource sharing based on that. 
Our concept builds on a simplified version of core-stateless resource sharing,
where we only use a few Drop Precedences.

\section{Packet level behavior}

In this section, we extend the Two-Rate, Three-Color Marker (trTCM) to provide
fairness on several timescales and show how we apply core stateless scheduling on the
marked packets.

\subsection{Packet Marking}
\label{ss:packetmarking}

\begin{figure}[h]
		\centerline{\includegraphics[width=0.40\textwidth]{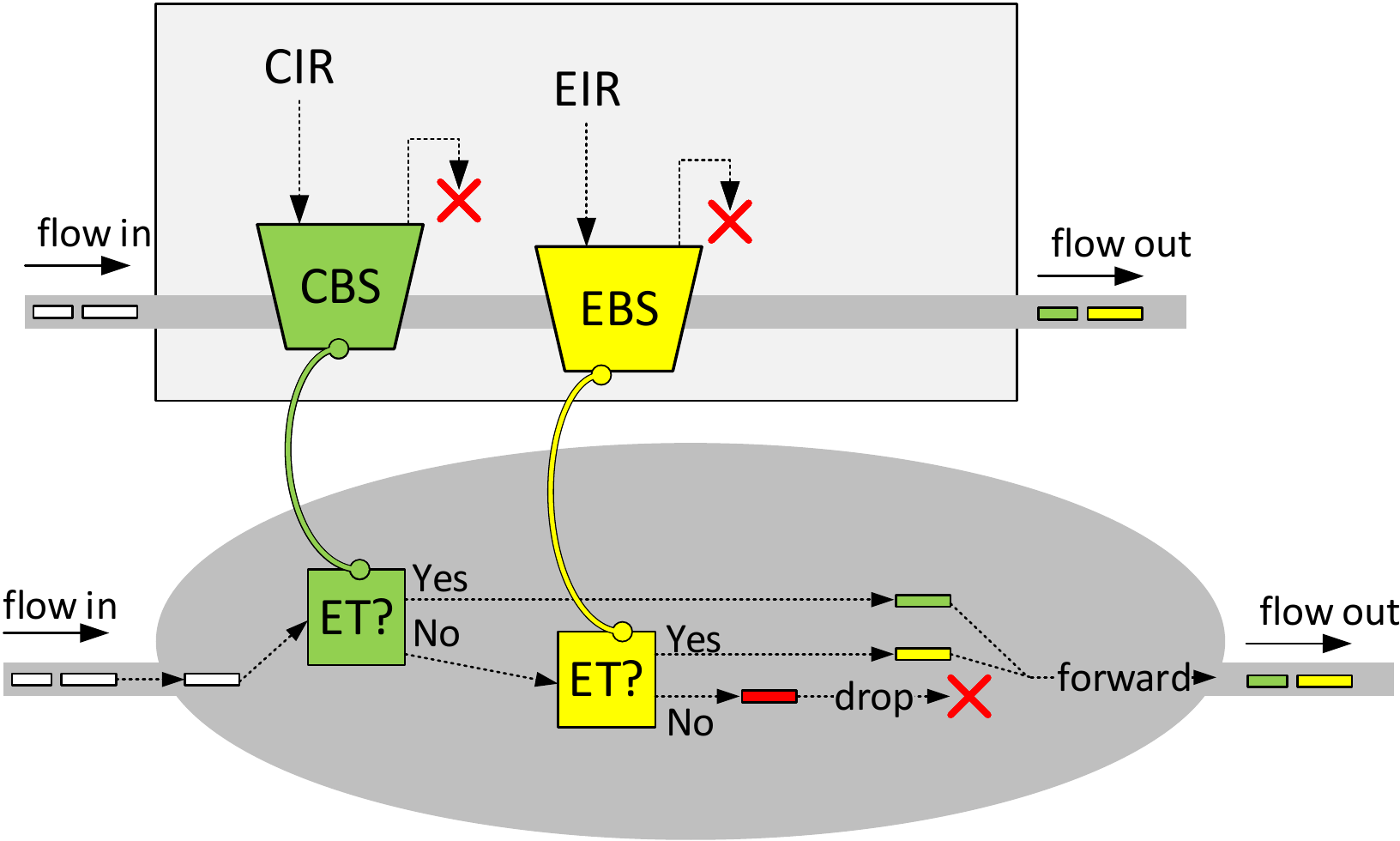}}
		\caption{trTCM bandwidth profile for a single priority 
	\label{fig:mefbwp}}
		\centerline{\includegraphics[width=0.48\textwidth]{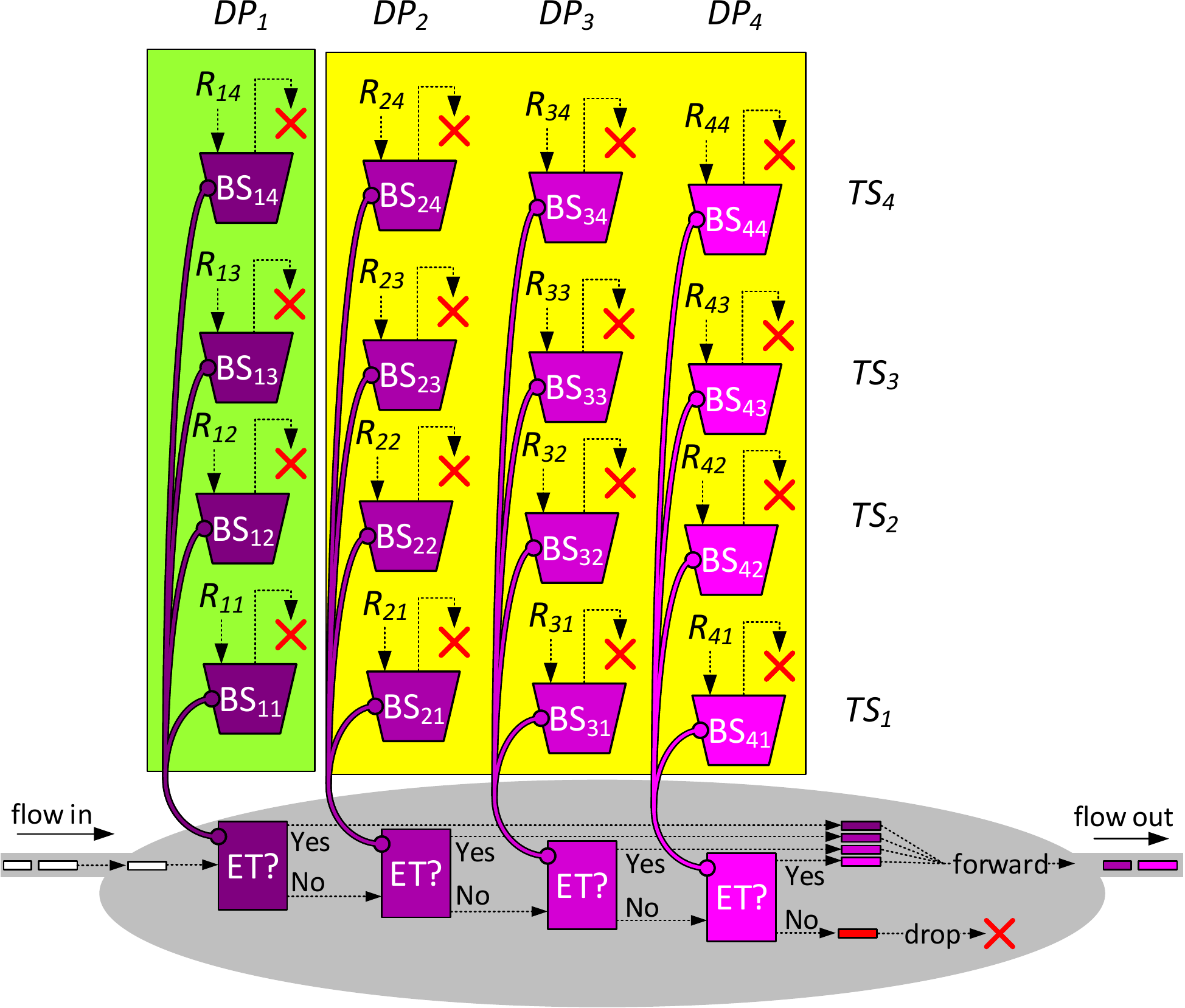}}
\caption{A $4 \times 4$ Multi Timescale Bandwidth Profile
	\label{fig:44bwp}}
\end{figure}

MEF currently uses flavors of trTCM for bandwidth profiling \cite{MEF62}. 
The simplest case, 
when a single priority is used and there is no token sharing or color awareness, 
is depicted on Fig. \ref{fig:mefbwp}.
It has two rate parameters, the guaranteed Committed Information Rate
(CIR) and the non-guaranteed Excess Information Rate (EIR).
Both has an associated token bucket  (Committed Burst Size (CBS) and Excess
Burst Size (EBS)), 
whose sizes are typically set to $CBS \approx  CIR \times RTT$ and $EBS \approx
EIR \times RTT$, where $RTT$ denotes the round trip time.
A packet is marked green (conform to CIR), yellow (conform to EIR) or dropped
(red) based on the amount of tokens in the associated token buckets. 
A bucket must contain at least as many tokens as the packet size, marked as
Enough Tokens (ET?).
If a packet is marked a given color, that many tokens are removed from the
relevant bucket.

We extend the trTCM by increasing the number of colors (i.e.\ drop precedences
(DPs)) and by introducing multiple token buckets per drop precedence,
representing different timescales (TSs).
An example for this Multi Timescale Bandwidth Profile (MTS-BWP) is shown on Fig
\ref{fig:44bwp}, where the number of DPs is $N_{DP}=4$ and the number of TSs is
$N_{TS}=4$.

The darkest purple color (DP 1) is similar to green in the sense that we intend
to guarantee transmission of packets marked dark purple. 
The lighter colors are similar to yellow, though we intend to provide a more
refined service for them than simple non-guaranteed delivery.
The token rate of the bucket associated with drop precedence $dp$ and timescale
$ts$ is $R_{dp,ts}$ and the bucket size of BS$_{dp,ts}$ is set to $BS_{dp,ts} \approx R_{dp,ts} \times TS_{ts}$. 
An example time scale vector with $N_{TS}=4$ is $TS=\{TS_1, \ldots,
TS_4\}=\{0.1, 1, 10, 100~\textrm{sec}\}$, where we assume that $TS_1$ is
in the order of magnitude of the typical RTT.

That is, $\mx{R}=\{R_{dp,ts}\}$ and $\mx{BS}=\{BS_{dp,ts}\}$ are matrices of size $N_{DP} \times N_{TS}$.
A packet can be marked a given DP value $dp$ if all buckets BS$_{dp,ts},\ \forall ts\in\{1,\ldots,N_{TS}\}$ (which we shorthand as $\forall ts$) contain enough tokens. 
Upon successful marking all respective buckets are decreased with the packet
size.

If we want to enable more bursty traffic on lower timescales, we have to 
offer smaller bandwidth on higher timescales. 
Thus the rows of $\mx{R}$ are monotonically decreasing, i.e. 
\begin{equation}
\label{eq:monoton}
R_{dp,ts+1}\geq R_{dp,ts},\ \forall dp,ts.
\end{equation}

\subsection{Active Queue Management (AQM) algorithm}

We assume a FIFO buffer with {\it drop the largest DP from head} AQM. 
More precisely when the buffer is full, we determine the largest DP which has
packet present in the buffer and drop the first packet from the head which has
this DP.

This algorithm can be implemented efficiently by applying the AQM in
\cite{SCDemo18} for 4 DPs.

This behavior can also be approximated on existing hardware by 
representing drop precedences with DSCPs; 
configuring these DSCPs to the same queue; 
and configuring DP specific TailDrop thresholds
to drop the different DPs at
increasing queue lengths (the largest DP at the smallest queue length)
\cite{rfc7806}.

\section{Fluid model of MTS-BWP}

We analyze the performance of MTS-BWP in a fast and simple fluid simulator,
because our focus is not the packet level behavior and the interaction of
congestion control and scheduling, as there is other work focusing on that,
e.g.\ \cite{ANRW18}.
Rather we are interested in how the newly introduced bandwidth profile can
provide fairness on several timescales, when the system is ideal, 
i.e.\  the used congestion control 
can utilize its share instantaneously 
and with no packet loss.

\subsection{High level model}
We model a common bottleneck shared among several {\it nodes} with identical
BWP configuration. 
A node may have one or several active {\it flows}. 
When no BWP is applied, nodes share the bottleneck proportionally to the number
flows within the nodes. 
The applied BWP constraints this allocation, all traffic with DP above {\it
congestion DP} is transmitted and with DP below is discarded.
Bandwidth allocation within the congestion DP is still 
proportional with the number of flows. 
DPs signify priority in the sense that whenever congestion occurs the transmission on higher DPs (i.e., traffic marked with higher DP value) is reduced first. 

The token level $TL^n_{dp,ts}(t)$ in bucket $\textrm{BS}_{dp,ts}$ of node $n$ is maintained and fluid flows out
from each bucket of a given DP according to the transmission rate of the node on
that DP $th_{dp,n}$.

Our ideal system  model (istantenous adaptation, no packet loss) assumes no
bottleneck buffer and 0 RTT. Consequently we set $BS_{dp,1}=0$ in the fluid
simulator (which results in maximum fluid rate of $R_{dp,1}$ on a given $dp$,
see Eq. \ref{eq:BDdpn}).

\subsection{Simulator model}
\label{s:simmodel}

\subsubsection{System parameters}
\label{ss:param}

The system is described by the following parameters:
\begin{itemize}
	\item $C$: the total service capacity;
	\item $N$: the number of nodes;
	\item $\mx{R}, \mx{BS}$ (assumed to be the same for all
	nodes);
	\item $f^{\max}$ flow limit: the maximum number of concurrent
	active flows at each node; further flows are discarded upon arrival. 
\end{itemize}

\subsubsection{Traffic Model}
\label{ss:inp}

We use a compound Poisson point process with a discrete file sizes distribution
(given by possible file sizes and associated probabilities) as an input, and
based on that we simulate the arrival time and the size of each arriving flow
for each node.

Each node has a \emph{nominal speed} $S_n$  
in all investigated situations, we will
stick to the natural choice of $S_n=C/N, \, n=1,\dots,N$. 
The \emph{nominal load} of each node can then be calculated as
$$\textrm{nominal load}=\textrm{average file size}\times \textrm{arrival rate}/ S_n.$$
The \emph{system load} is the average of the nominal loads for all nodes. 
The system is \emph{underloaded} if its load is less than 1, and it is \emph{overloaded}
otherwise.
A node has \emph{low load} if its nominal load is less than the system load, and
it has \emph{high load} otherwise.


\subsubsection{Discrete event simulator}
\label{ss:sim}

A discrete event simulator runs a simulation using a given traffic input (arrival time and file size series) for a given set of system parameters. The simulator identifies the following events:
\begin{enumerate}
	\item flow arrival,
	\item flow finishing,
	\item a token bucket emptying,
\end{enumerate}
and keeps track of the following values:
\begin{enumerate}
	\item simulation time;
	\item list of active flows;
	\item remaining size of each active flow;
	\item token bucket levels, $TL^n_{dp,ts}$. 
\end{enumerate}

These variables are sufficient to determine the time and type of the next event
and the current \bwa which applies until
the next event.
The simulator then proceeds to the next event, updates all information and iterates this loop.

Once finished, the simulator provides the following information as an \emph{output}:
\begin{enumerate}
	\item list of all event times (including flow arrivals and departures);
	\item list of the \bwa for each node and each DP between events;
	\item flow count for each node at each event time.
\end{enumerate}

\subsection{Bandwidth allocation model} 

At any given point in time, we collect the throughput bounds determined by the
current bucket levels for a drop precedence $dp$ at node $n$ into a $N_{DP}
\times N$ matrix, denoted by $\mx{\BD}$, whose elements are
\begin{equation}
\label{eq:BDdpn}
\min_{ts=1 \dots N_{TS}}\lbrace R_{dp,ts}: TL^n_{dp,ts}=0 \rbrace.
\end{equation}

To present the bandwidth allocation we need the following notation:
\begin{itemize}
\item $f_n$ is the number of flows in node $n$,
\item $th_n$ is the throughput of node $n$, initialized with 0,
\item $e_n$ shows that node $n$ is eligible for increase, initialized with True. 
\end{itemize}

The iterative algorithm to calculate the bandwidth allocation is as follows. 

The congestion DP is calculated as
$$dp_c=\min\Big\{i:\sum_{dp=1}^{i}\sum_{n=1}^{N} \BD_{dp,n}\geq C\Big\}.$$
$th_n$ is initialized for all $n$ as
$$th_n = \sum_{dp=1}^{dp_c-1}\BD_{dp,n}.$$
Then the procedure iterate the following 3 steps until $\sum_{n=1}^N th_n= C$:
\begin{enumerate}
	\item Nodes with $th_n= \sum_{dp=1}^{dp_c}\BD_{dp,n}$ are set to non-eligible ($e_n=$ False.)
	\item Mark all eligible nodes for which the ratio $th_n/f_n$ is minimal among all eligible nodes.
	\item Increase $th_n$ for all marked nodes by $f_n \cdot \delta$, where $\delta>0$ is calculated as the maximal possible increase such that the following remain valid:
	\begin{itemize}
		\item $th_n\leq \sum_{dp=1}^{dp_c}\BD_{dp,n}$ for all $n$,
		\item the ratio $th_n/f_n$ from among all marked nodes does not increase beyond the \emph{second smallest ratio} $th_n/f_n$ from among all eligible nodes, and
		\item $\sum_{n=1}^N th_n\leq C$.
	\end{itemize}
\end{enumerate}
From $th_n$ and $BD_{dp,n}$ calculating the per DP throughput $th_{dp,n}$ is
straightforward.

\section{Dimensioning guidelines}
\label{s:dim}

This section focuses on the dimensioning of the token rate matrix $\mx{R}$ and the
token bucket size matrix $\mx{BS}$ (defined in Section \ref{ss:packetmarking}). 
Proper dimensioning of $\mx{R}$ and $\mx{BS}$ are vital to obtain the desired properties
of the bandwidth profile. 
We consider a system with $N$ nodes with identical MTS-BWP configuration over a
bottleneck link with capacity $C$. The required properties are the following:
\begin{enumerate}
	\item\label{reqfirst} there are $N_{\fs}$ predefined download speeds
	$BW_1,BW_2,\dots,BW_{N_{\fs}}$ (decreasing) provided for files of sizes
	$\fs_1,\fs_2,\dots,\fs_{N_{\fs}}$ (increasing) arriving at a previously
	inactive node ($BW_1$ is also the peak rate provided to a node after an
	inactive period);
	\item provide the nominal speed $S_n=C/N$ to each node in long-term average;
	\item ensure $BW_{N_{\fs}}>S_n$;
	\item provide the minimum guaranteed speed
	$G_1,G_2,\dots,G_{N_{TS}}$ (decreasing) in case of $TL^n_{11} =0 , TL^n_{12} =0
	,\dots, TL^n_{N_{TS} =0}$ respectively;
	\item\label{reqlast} guarantee work conserving property (i.e.\ when there is
	traffic, the full link capacity shall be used).
\end{enumerate}

In the following analysis we focus on the $N_{DP}=4$ and $N_{TS}=4$ case, 
which allows for $N_{\fs}=N_{TS}-1=3$ file sizes with predefined downloads speeds, 
but it is straightforward to generalize for $N_{TS}>4$. 
We aim to minimize $N_{DP}$ and will settle at $N_{DP}=4$, providing insight
to how the 4 DPs are used as well as what happens for fewer DPs.

\subsection{Token rate matrix $\mx{R}$}
\label{ss:dimR}

In this part we present a simple dimensioning method for a $4\times 4$ matrix $\mx{R}$ based
on the requirements \ref{reqfirst}--\ref{reqlast} above. All rows of $\mx{R}$ should be decreasing according to \eqref{eq:monoton}.

We use the following intuitive guidelines for $\mx{R}$:
\begin{itemize}
	\item DP 1 is used for the guaranteed speeds $G_1, G_2, \dots$ and not
	intended to be the limiting DP;
	\item DP 1 and 2 are used to be able to reach the predefined download speeds
	($BW_1,BW_2,\dots$) by a low load node in situations when most nodes are on $ts=N_{TS}$;
	\item DP 3 or 4 is the congestion DP for high load nodes while low load nodes are inactive;
	\item DP 4 is used to guarantee the work conserving property.
\end{itemize}

In accordance with these guidelines, we propose the structure
\begin{align}
\label{eq:R}
\mx{R}\!=\!\!
\begin{bmatrix}
G_1 & G_2 & G_3 & G_4\\
BW_1\!-\!G_1 & BW_2\!-\!G_2 & BW_3\!-\!G_3  & \frac{C-BW_1}{N-1}\!-\!G_4\\
* & * & * & S_n\!-\!\frac{C-BW_1}{N-1}\\
C & C & C & C
\end{bmatrix}\nonumber\\
\end{align}

The first row (DP 1) is straightforward and simply implements the guaranteed
speeds. Note that $G_1\leq S_n$ needs to hold to avoid congestion on DP 1.

$R_{2,1}$ is calculated so that $R_{1,1}+R_{2,1}=BW_1$ to ensure the predefined
download speed $BW_1$ on DP 1 and 2; similarly, $R_{2,2}=BW_2-R_{2,1}$ and
$R_{2,3}=BW_3-R_{3,1}$. 

Our next remark is that $R_{3,4}$ is defined so that
\begin{align}
\label{eq:rr}
R_{1,4}+R_{2,4}+R_{3,4}=S_n
\end{align}
holds; this important property will be called the \emph{return rule}. 
First note that if $R_{1,4}+R_{2,4}+R_{3,4}\leq S_n$, then any node $n$ with
nominal load larger than 1 will continue to deplete their token buckets and
eventually end up with
\begin{align}
\label{eq:badhist}
TL^n_{1,N_{TS}} = TL^n_{2,N_{TS}} = TL^n_{3,N_{TS}} = 0.
\end{align}
Actually, \eqref{eq:badhist} is exactly the way bad history is described within the system.

The return rule \eqref{eq:rr} provides two important guarantees: in long-term average, only bandwidth $S_n$ is guaranteed \emph{on DP1--DP3} for any node $n$, but since $S_n=C/N$, this also means that no node will be ``suppressed'' in long-term average by the other nodes.

Also, over a time period when all other nodes $n$ are either inactive or have
bad history as in \eqref{eq:badhist}, any node $\bar n$ with nominal load less
than $1$ will eventually have $TL^{\bar n}_{3,4}>0$, and thus potentially have
access to a rate larger than $S_{\bar n}$ (the node returns from ``bad history'' to
``good history'', hence the name of the rule). 
The general form of the return rule would be that there exists a $dp_r$ such
that $\sum_{dp=1}^{dp_r} R_{dp,N_{TS}}=S_n$.

Next up is $R_{2,4}$, which is defined so that
$$(N-1)(R_{1,4}+R_{2,4})+BW_1=C.$$
This will ensure that in the case when a single$  $ node becomes active while all other
nodes are either inactive or have bad history as in \eqref{eq:badhist}, 
the congestion DP will change to 2, with the single active node having rate
$BW_1$ and other nodes having rate $(R_{1,4}+R_{2,4})$ allocated. 

The last row guarantees the work conserving property: as long as at least one node is active, it has access to the entire capacity $C$.

Finally, the system is relatively insensitive to the exact values of the
elements marked with an $*$ since typically other elements will limit the
bandwidth: high load nodes have a bad history and thus are limited by
$R_{.,4}$, while targets for low load nodes are realized on DPs 1 and 2 and are thus
limited by $(R_{1,.})$ and $(R_{2,.})$. 
Elements marked with an * can be selected arbitrarily as long as row 3 of
$\mx{R}$ is decreasing.

Further remarks:
The file sizes for the predefined download speeds only affect the bucket size
matrix $BS$, detailed in the next subsection.
For typical choices of the parameters, the rows of $\mx{R}$ are
monotonically decreasing, but in case they are not, $\mx{R}$ needs to be
adjusted, which we neglect here.
More (or fewer) timescales can be introduced in a straightforward manner to accommodate more (or fewer) predefined file sizes and download rates.
In case of fewer DPs, we need to choose:
\begin{itemize}
\item Omitting the first row of $\mx{R}$ results in no strictly guaranteed
speeds.
\item Omitting the second row removes the predefined download speeds, resulting in a system very similar to trTCM, with nearly no memory.
\item Omitting the third row violates the return rule.
\item Omitting the last row  results in a non-work-conserving system, where it may occur that their previous history limits nodes to the point where less than the available capacity is used.
\end{itemize}

\begin{exmp}
\label{ex:1} For the parameters $N=5$, $C=10$ (Gbps), guaranteed speeds $G_1=G_2=G_3=2, G_4=0.75$ (Gbps), file sizes are $\fs_1=0.1, \fs_2=1, \fs_3=11.25$ (GByte) and download speeds $BW_1=6, BW_2=4, BW_3=3$ (Gbps), the following $4\times 4$ matrix is suitable:
\begin{equation}
\label{eq:Rnum}
\mx{R}=
 \begin{bmatrix}
 	2 & 2 & 2 & 0.75\\
 	4 & 2 & 1 & 0.25\\
 	10 & 10 & 1 & 1\\
 	10 & 10 & 10 & 10
\end{bmatrix}
\end{equation}

\end{exmp}

\subsection{Bucket size matrix $\mx{BS}$}

The sizes of the buckets are calculated from the rates in $R$ and the list of timescales $TS$, which we define as
\begin{align}
\label{eq:ts}
TS=[0,\fs_1/BW_1,\fs_2/BW_2,\fs_3/BW_3].
\end{align}

$TS_1=0$ represents the ideal behavior of the fluid model.
The remaining timescales correspond to download times of the
predefined file sizes. We use the last timescale ($TS_4$) to define how long a
node must send with bandwidth at least $S_n$ to be considered to have bad history.
In Example \ref{ex:1}, we actually set $TS_4 = 30$ sec (to allow a 30 second active period, before a node is
considered to have bad history) and calculate $\fs_3$ accordingly.


We set the bucket sizes according to the formula
\begin{align}\label{eq:bs_ij}
&BS_{dp,ts}= \\
&\left\lbrace\begin{array}{lrl}
0 \hfill & \mbox{for }ts=1\\
TS_{2}(R_{dp,1}-R_{dp,2})& \mbox{for } ts=2\\
\sum_{k=2}^{ts}(TS_{k}-TS_{k-1})(R_{dp,k-1}-R_{dp,ts}) &\mbox{for } ts>2
\end{array}\right.
\nonumber
\end{align}
which will result in a previously inactive node emptying bucket
$\textrm{BS}_{dp,ts}$ after time $TS_{ts}$ (assuming the rate at DP $dp$ is
limited only by the node's own history and not by other nodes), taking into
account the fact that it uses different bandwidth on different timescales.
Buckets with $BS_{dp,ts}=0$ act as rate limiters in the fluid model, due
to Eq. \ref{eq:BDdpn}.

Note that when using the above $\mx{R}$ and $\mx{BS}$ dimensioning method, the
flow throughput of a single flow of size $\fs_2$ can reach as high as 
\begin{equation}
\label{eq:BW2star}
BW'_2 = \frac{TS_2 \cdot BW_1 + (TS_3-TS_2) \cdot BW_2}{TS_3} \geq BW_2. 
\end{equation}
If one wants to replace the current {\it maintainable download speed
	requirement} to a {\it flow throughput requirement} for $\fs_2$, $BW_2$ in
\eqref{eq:R} should be replaced by the (slightly smaller) solution of Eq.
\eqref{eq:BW2star} for $BW_2$  when setting the left-hand side equal to the {\it
	flow throughput requirement}. 
For Example \ref{ex:1}, and for the anticipated meaningful input values, the
difference between $BW_2$ and $BW'_2$ is very small; specifically, $BW_2'=4.1333$ (Gbps). 

Also note that the above calculations are for the fluid model; for actual packet-based networks, bucket sizes $BS^*_{dp,ts}$ must have a minimum: at least MTU (maximum transmission unit) to be able to pass packets, and they must also be able to allow bursts on the RTT timescale. 
In summary,
$$BS^*_{dp,ts}=\max(BS_{dp,ts},\, MTU,\, R_{dp,ts} \cdot RTT).$$

\section{Simulation}
\label{s:res}


\subsection{Simulation parameters}

In all simulations, the MTS-BWP rates ($\mx{R}$), bucket sizes $\mx{BS}$
and the system parameters are set according to Example \ref{ex:1}.
In the input process, we use the file sizes $\fs_1$ and $\fs_2$ from
Example \ref{ex:1} with identical $50\%$ probability. $f^{\max}=20$  is for each
node.

We have two groups of nodes with identical nominal loads
within a group. We specify the nominal load for low load nodes ({\it low load})
and the {\it system load}, and calculate the nominal load for high load nodes
using the equations in Section \ref{ss:inp}.
The simulation setups are summarized in Table \ref{t:setups}, 
with the number and load of each node type varying for a total of $2\times
10+2\times 6$ actual setups.
\setlength\extrarowheight{1pt}
\begin{table}
	\centering
	\begin{tabular}{|c|c|c|c|c|}
		\hline
		Setup & $N_\textrm{low}$ & $N_\textrm{high}$ & low load & system load\\
		\hline
		{A} & 1 & 4 & \multirow{2}{*}{0.5} & 0.6, 0.7, 0.8, 0.9, 0.95,\\
		\cline{1-3}
		B & 2 & 3 & & 1.0, 1.1, 1.2, 1.5, 2.0 \\
		\hline		
		C & 1 & 4 & 0.5, 0.6, 0.7, & \multirow{2}{*}{1.1}\\ 
		\cline{1-3}
		D & 2 & 3 & 0.8, 0.9, 0.95 & \\
		\hline
	\end{tabular}
	\caption{Simulation setups and parameters}
	\label{t:setups}
\end{table}

\subsection{Example simulation}

\begin{figure}
	\centerline{\includegraphics[width=0.46\textwidth]{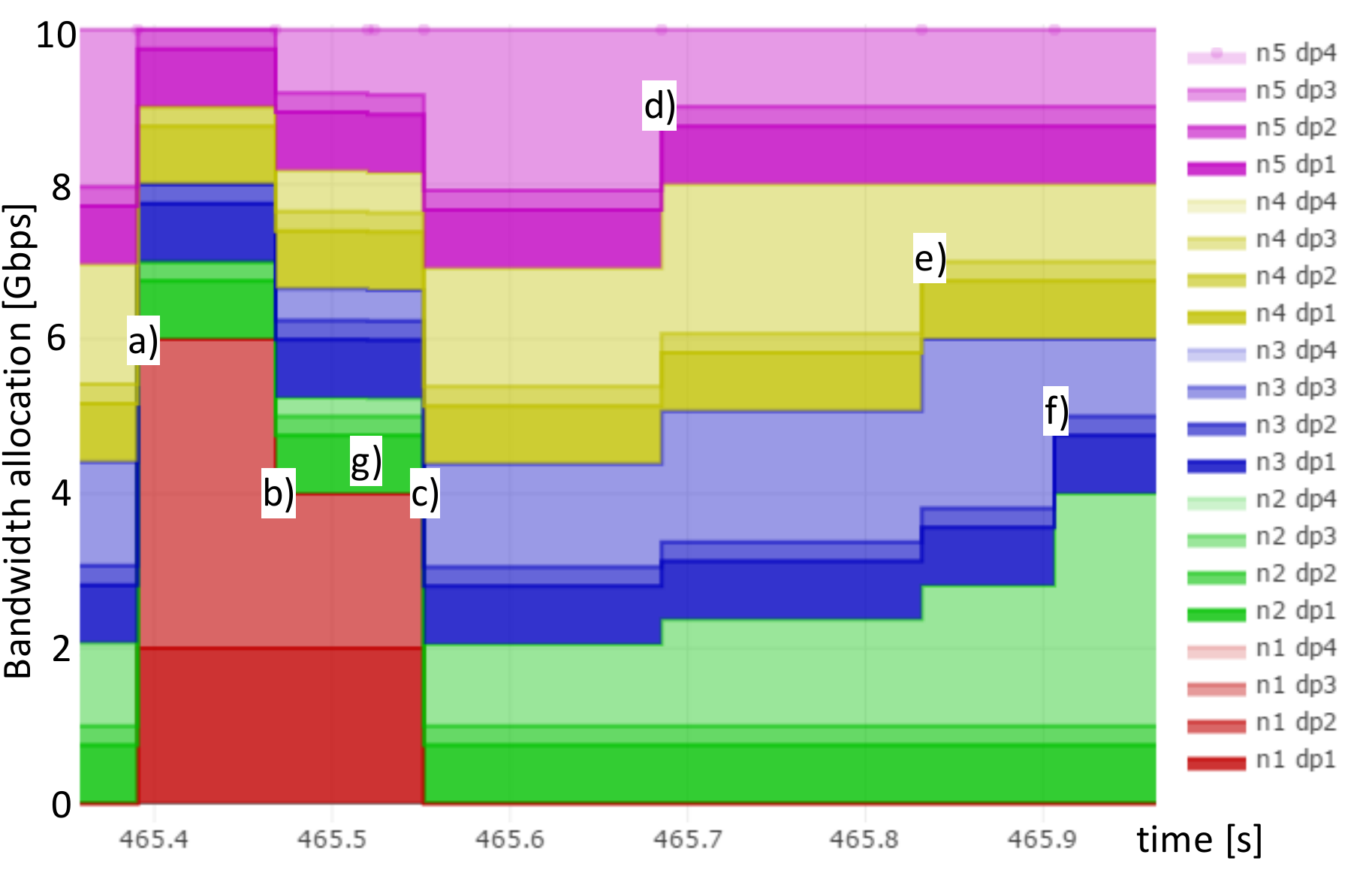}}
	\caption{ Example of bandwidth allocation over time
		\label{fig:timeseries}}
\end{figure}

Figure \ref{fig:timeseries} depicts the evolution of the bandwidth
allocation in a time interval for setup A with system load of $0.95$.
Colors correspond to nodes and shades within a color correspond to DPs. 
Node 1 (red) is the low load node. Some events are also marked (a)--(g).

Node 1 is inactive in the beginning, and the congestion DP is 3.
Then a large flow starts in node 1 (a) and the
congestion DP changes to 2.
Node 1 starts using 2 Gbps ($R_{1,1}$) + 4 Gbps ($R_{2,1}$) of the
available capacity on DP 1 and 2 respectively, 
while nodes 2--5 start using 0.25 Gbps ($R_{1,4}$)
+ 0.75 Gbps ($R_{2,4}$) respectively.
($R_{1,1}$ and $R_{2,1}$ was dimensioned for exactly this case; while the congestion DP is 2, all traffic on DP 2 can be transmitted.)

As time progresses, the buckets $\textrm{BS}_{2,2}$ and $\textrm{BS}_{1,2}$ of node 1
becomes empty ({b}),
and the bandwidth share of node 1 drops accordingly to 2 Gbps ($R_{1,2}$) + 2
Gbps ($R_{2,2}$). 
The congestion DP switches back to 3, but DP 3 is dominated by nodes 2--5,
because those nodes have high numbers of flows, while node 1 has only a single
flow.
That single large flow can still achieve $BW_2$ throughput as dimensioned.

Once node 1 finishes its flow ({c}), the
available bandwidth is reallocated to nodes 2--5 on DP 3.
Then, buckets which were filled previously (specifically BS$_{3,4}$) of nodes
2--5 empty one by one, and their bandwidth shares on DP 3 drop accordingly: 
first for node 5 ({d}),
then node 4 (e), 
then node 3 (f). 
The exact order depends on the bucket levels of $B_{3,4}$ of each node, which
depend on their earlier history, not visible in the example time interval.

In the meantime, new flow arrivals and finished services at nodes 2--5 may
occur and cause minor changes in the bandwidth allocation, e.g.\ a flow at node
2 finishes at ({g}).

\subsection{Statistical results}
\label{ss:statres}

Based on the simulator output, we calculate the following two statistics:
the node bandwidth for active periods (periods when
there is no flow at the respective node are excluded); and
the flow bandwidth for the different flow sizes, which is the flow size divided
by the download time.
For both we plot average, 10\% best and 10\% worst cases, the error bars displaying
the worst $10\%$--best $10\%$ interval, with a dot marking the average. 
Averaging and percentiles are weighted according to time for node throughputs, while they are weighted according to the number of flows for flow throughputs.
All statistics are evaluated for a 1-hour run (with an extra initial warm-up
period excluded from the statistics).

We compare the suggested MTS-BWP with matrix $\mx{R}$ versus trTCM profile
(CIR=$S_n$=2~Gbps, EIR=$C-S_n$=8~Gbps) as baseline for various setups.

\begin{figure*}[t]
	\centerline{\includegraphics[width=0.90\textwidth]{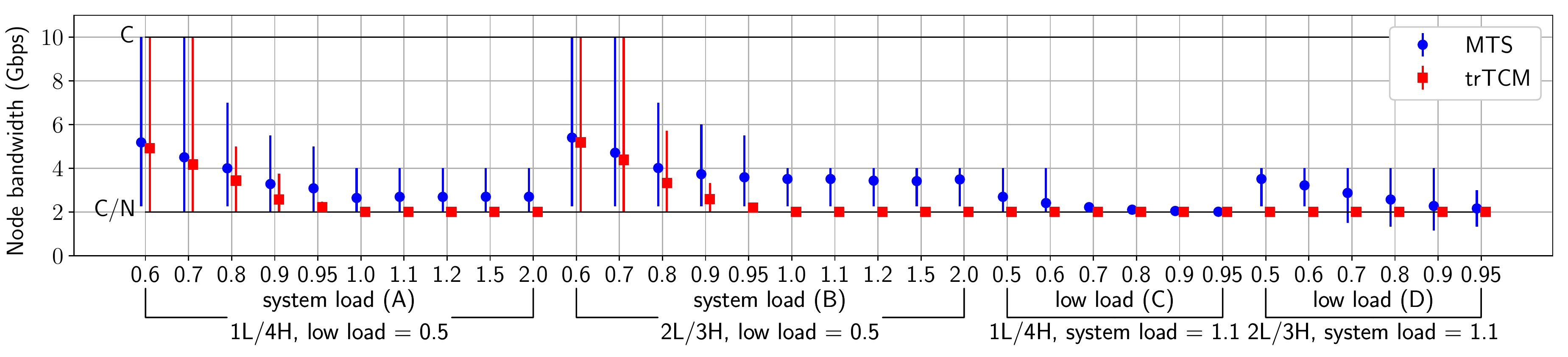}}
	\caption{Node bandwidth for low load nodes for trTCM vs. MTS
	BWP}\label{fig:44mefl}
	\centerline{\includegraphics[width=0.90\textwidth]{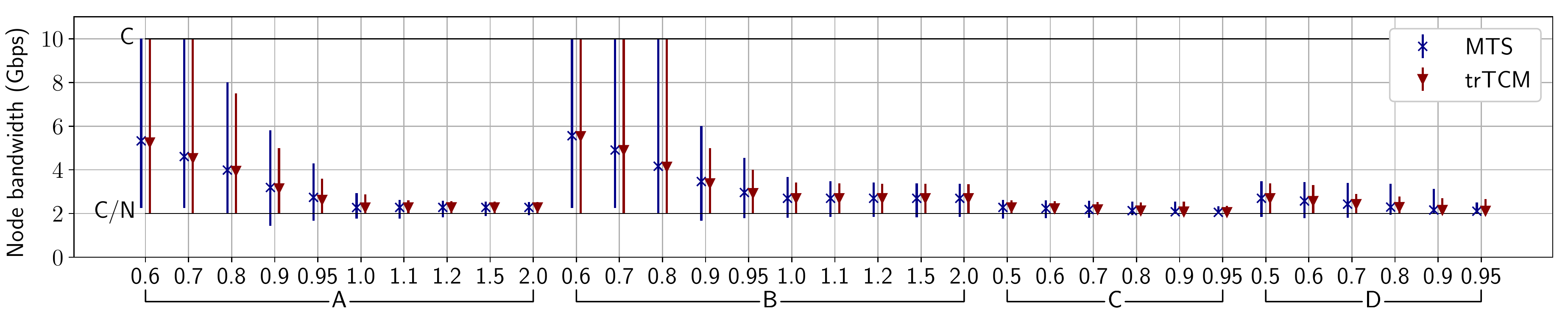}}
	\caption{Node bandwidth for high load nodes for trTCM vs. MTS BWP}\label{fig:44mefh}
	\centerline{\includegraphics[width=0.90\textwidth]{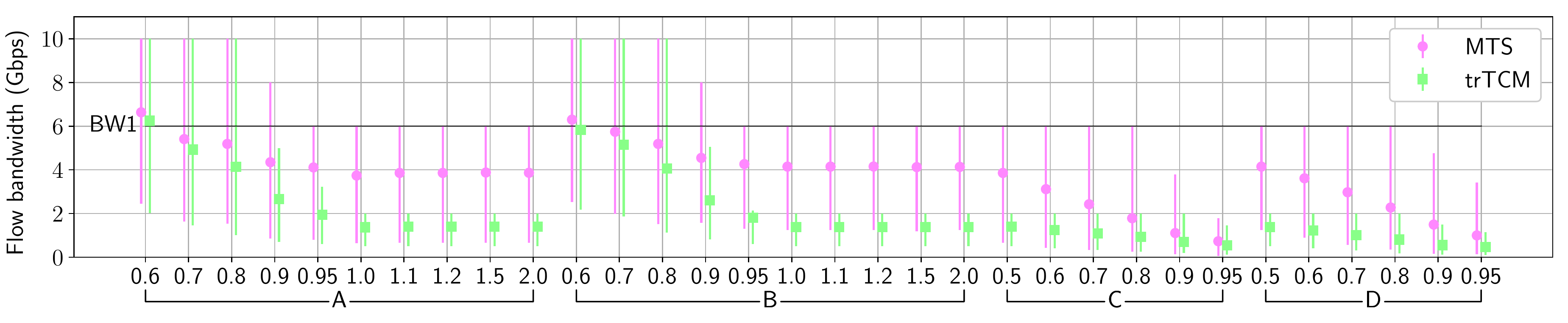}}
	\caption{Flow bandwidth for small flows (100 MB) at low load nodes for trTCM vs. MTS BWP}\label{fig:44meflfs1}
	\centerline{\includegraphics[width=0.90\textwidth]{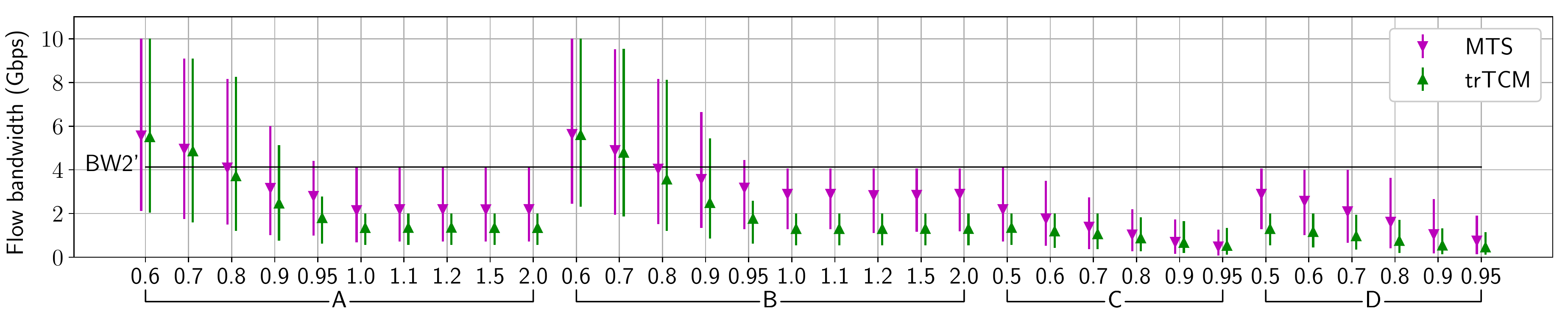}}
	\caption{Flow bandwidth for large flows (1 GB) at low load nodes for trTCM vs. MTS BWP}\label{fig:44meflfs2}
	\centerline{\includegraphics[width=0.90\textwidth]{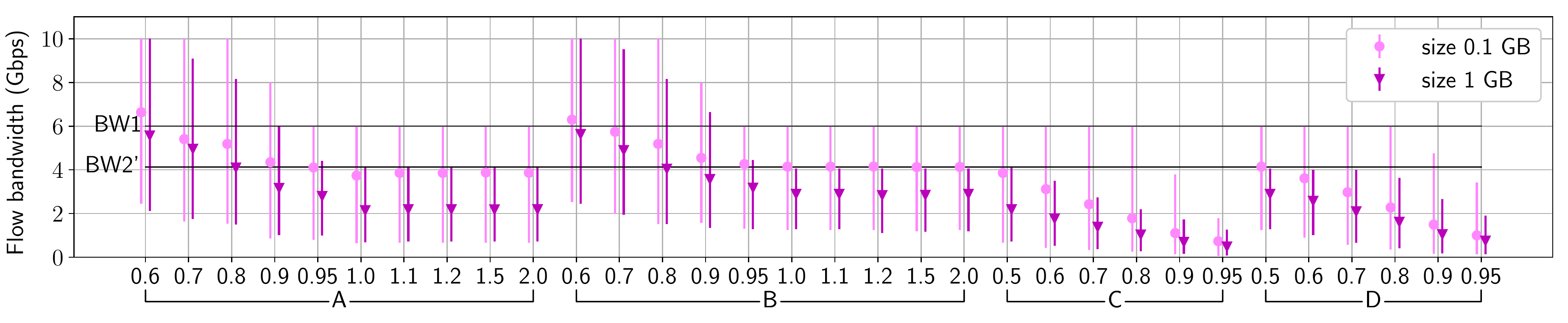}}
	\caption{Flow bandwidth for small flows vs. large flows at low load nodes for MTS BWP}\label{fig:44lfs1fs2}
	\centerline{\includegraphics[width=0.90\textwidth]{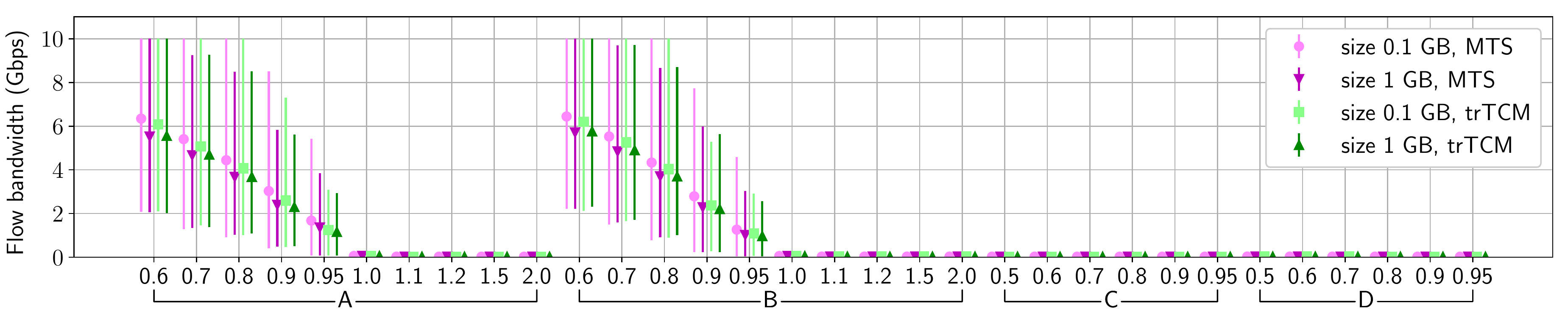}}
	\caption{Flow bandwidth for small vs. large flows at high load nodes for trTCM vs. MTS BWP}\label{fig:44mefhfs1fs2}
\end{figure*}

Figure \ref{fig:44mefl} displays node bandwidth statistics for low load nodes
for trTCM vs. MTS.
MTS consistently outperforms trTCM in allocating more bandwidth to low load
nodes. 
The average bandwidth for MTS is higher in every scenario, and the best
possible case (best $10\%$ values) is also significantly higher in most scenarios. 
MTS provides the most improvement in scenarios where the system is overloaded,
for small system loads trTCM also performs well. 

Low load node(s) perform better in the 2L/3H setup than in the 1L/4H
setup, because $\mx{R}$ protects a low load node better from 3 high load nodes
than from 4.
Finally, as the load of the low load node approaches 1, the difference
between trTCM and MTS gradually disappears.

Figure \ref{fig:44mefh} displays the same statistics for high load nodes. 
The most important observation here is that the considerable gain for low load
nodes in Figure \ref{fig:44mefl} comes at virtually no cost to high load nodes:
the difference between the average bandwidth for high load nodes for trTCM vs.
MTS BWP is negligible.
The reason is that while traffic from low load nodes is indeed served faster,
but the total amount of traffic served from low load nodes is the same.
This means that for high load nodes, which are active longer, the effect on node
bandwidth is negligible, especially for the average. (It matters little whether we
decrease the same amount of bytes in a big burst or for a longer period with
smaller bandwidth.)

Next we examine the prioritization of small flows ($\fs_1$) vs. large flows
($\fs_2$) provided by MTS-BWP compared to trTCM BWP.
Figure \ref{fig:44meflfs1} shows flow bandwidth statistics for small flows in
low load nodes.
MTS outperforms trTCM in allocating more bandwidth in these cases 
for every setup, but particularly for overloaded systems, where the
difference is huge, both for average and also for best $10\%$ values. 
Also, as the low load is approaching 1, the difference between trTCM and MTS
diminishes (just as for the node bandwidth, see Figure \ref{fig:44mefl}), but
that is as it should be. 
Also, for MTS BWP, the best $10\%$ values for small flows reach $BW_1$ for
all scenarios where the low load is below $0.9$.

Figure \ref{fig:44meflfs2} displays the same statistics for large flows (1 GB)
at low load nodes. 
Again,  MTS outperforms trTCM significantly. 
The best $10\%$ throughput matches $BW_2'$ and is close to the dimensioned
$BW_2$ (see Section \ref{ss:dimR}). 

Figure \ref{fig:44lfs1fs2} compares flow bandwidth statistics for small vs.
large flows at low load nodes for MTS-BWP. 
It can be seen that small flows are successfully prioritized in all cases.

Finally, Figure \ref{fig:44mefhfs1fs2} displays flow bandwidth at high load
nodes for both small and large flows; and for both policies. 
There is a sharp distinction between underloaded systems and overloaded
systems: for underloaded systems, even at high load nodes, 
typically there are only very few active flows at the same time, 
resulting in relatively large flow bandwidths. 
However, as the system load approaches 1, the per flow bandwidth drops
gradually,
and for overloaded systems, the number of flows in high load nodes is always
close to the limit $f^{\max}=20$. 
Thus the flow bandwidth in high load nodes is typically close to $S_n/f^{\max}$, which
is very sensitive to the parameter $f^{\max}$. For these nodes, we consider the node
bandwidth to be a more informative statistics.

\section{Conclusion}

We have shown that the proposed Multi Timescale Bandwidth Profile can extend
fairness from a single timescale to several timescales. 
We provided a dimensioning method to deploy Service Level Agreements based on
MTS-BWP, 
which can provide target throughputs for a group of nodes.
The presented tool can differentiate between nodes using the same
service depending on their characteristics. 

Our simulation results showed the differences in network throughput for low
load and high load nodes. 
There were high throughput gains on low load nodes with marginal or no
throughput decrease on high load ones.

\bibliographystyle{IEEEtran}
\bibliography{bib}

\end{document}